\documentclass[10pt,aps,prl,twocolumn,superscriptaddress,preprintnumbers,floatfix,notitlepage]{revtex4-1}
\usepackage{hyperref}
\hypersetup{
    colorlinks=true,
    linkcolor=blue,
    filecolor=magenta,      
    urlcolor=blue,
}
\usepackage{amsmath,amsfonts,amssymb}
\usepackage{fancyhdr}
\usepackage{graphicx}
\usepackage{xspace}
\usepackage{rotating}
\usepackage[normalem]{ulem}
\usepackage{braket}
\usepackage{verbatim}
\usepackage{xcolor}
\usepackage{hyperref}
\usepackage[utf8]{inputenc}

\def\lsim{\mathrel{\raise.3ex\hbox{$<$\kern-.75em\lower1ex\hbox{$\sim$}}}}
\def\gsim{\mathrel{\raise.3ex\hbox{$>$\kern-.75em\lower1ex\hbox{$\sim$}}}}

\newcommand{\be}{\begin{equation}}
\newcommand{\ee}{\end{equation}}
\newcommand{\bea}{\begin{equation}\begin{aligned}}
\newcommand{\eea}{\end{aligned}\end{equation}}

\begin{document}

\title{Cosmic Superstrings Revisited in Light of NANOGrav 15-Year Data}

\author{John~Ellis}
\email[]{john.ellis@cern.ch}
\affiliation{Kings College London, Strand, London, WC2R 2LS, United Kingdom}
\affiliation{Theoretical Physics Department, CERN, Geneva, Switzerland}

\author{Marek~Lewicki}
\email[]{marek.lewicki@fuw.edu.pl}
\affiliation{Faculty of Physics, University of Warsaw ul.\ Pasteura 5, 02-093 Warsaw, Poland}

\author{Chunshan~Lin}
\email[]{chunshan.lin@uj.edu.pl}
\affiliation{Faculty of Physics, Astronomy and Applied Computer Science, Jagiellonian University,
30-348 Krak\'ow, Poland}

\author{Ville Vaskonen}
\email[]{ville.vaskonen@pd.infn.it}
\affiliation{Keemilise ja Bioloogilise F\"u\"usika Instituut, R\"avala pst. 10, 10143 Tallinn, Estonia}
\affiliation{Dipartimento di Fisica e Astronomia, Universit\`a degli Studi di Padova, Via Marzolo 8, 35131 Padova, Italy}
\affiliation{Istituto Nazionale di Fisica Nucleare, Sezione di Padova, Via Marzolo 8, 35131 Padova, Italy}

\begin{abstract}
We analyze cosmic superstring models in light of NANOGrav 15-year pulsar timing data. A good fit is found for a string tension $G \mu \sim 10^{-12} - 10^{-11}$ and a string intercommutation probability $p \sim 10^{-3} - 10^{-1}$. Extrapolation to higher frequencies assuming standard Big Bang cosmology is compatible at the 68\% CL with the current LIGO/Virgo/KAGRA upper limit on a  stochastic gravitational wave background in the 10 to 100 Hz range. The superstring interpretation of the NANOGrav data would be robustly testable by future experiments even in modified cosmological scenarios.\\
~~\\
KCL-PH-TH/2023-38, CERN-TH-2023-126, AION-REPORT/2023-07
\end{abstract}

\maketitle

{\bf Introduction:} The NANOGrav and other Pulsar Timing Array (PTA) Collaborations have recently presented evidence for the Hellings-Downs angular correlation in the common-spectrum process that had been observed previously by PTAs~\cite{NANOGrav:2023gor,Antoniadis:2023ott,Reardon:2023gzh,Xu:2023wog,NANOGrav:2023hde,NANOGrav:2023hvm,NANOGrav:2023hfp,Antoniadis:2023lym,Antoniadis:2023aac,Antoniadis:2023xlr,Reardon:2023zen,Zic:2023gta}. This discovery demonstrates that the common-spectrum process is due to gravitational waves (GWs) in the nHz range. However, their origin has not been established, and it is important to know how  alternative scenarios can be distinguished. The default scenario would be astrophysical, specifically that these GWs are emitted by supermassive black hole (SMBH) binaries. However, at this stage, it is not possible to exclude more exotic scenarios involving physics beyond the Standard Model, such as cosmic strings or a cosmological phase transition~\cite{NANOGrav:2023hvm}.

A cosmic string model was able to fit the NANOGrav 12.5-year data~\cite{Arzoumanian:2020vkk} that contained the first reported evidence for a common spectrum process in the nHz range~\cite{Ellis:2020ena,Blasi:2020mfx,Blanco-Pillado:2021ygr} that was subsequently confirmed by other PTAs~\cite{Goncharov:2021oub,Chen:2021rqp,Antoniadis:2022pcn}. In this paper we revisit cosmic strings in light of the NANOGrav 15-year data, discussing modifications of the previous model that enable it to accommodate the updated NANOGrav data as well as other experimental constraints. A key prediction of the cosmic string model is the presence of a stochastic gravitational wave background (SGWB) that extends over many orders of magnitude in frequency, whereas the extension of the SMBH binary model to higher frequencies is subject to larger uncertainties~\cite{Ellis:2023owy}. Our aim in this paper, therefore, is to explore the robustness of the cosmic string predictions and whether it can avoid making testable predictions for higher-frequency GW detectors.

{\bf Cosmic (Super) Strings:} Cosmic strings are a common feature of numerous extensions of the Standard Model~\cite{Jeannerot:2003qv,King:2020hyd}, usually realised through the spontaneous breaking of a U(1) symmetry in the early Universe~\cite{Nielsen:1973cs}. As discussed in more detail below, the motions and intercommutations of a network of cosmic strings generate a spectrum of GWs extending over many orders of magnitude in frequency whose magnitude and shape are determined by a single parameter, the string tension $G \mu$. This scenario was analysed in the context of the NANOGrav 12.5-year data~\cite{Ellis:2020ena,Blasi:2020mfx,Blanco-Pillado:2021ygr}, and provided a very good fit to the observations. However, it is immediately apparent from inspection (for example) of Fig.~2 in~\cite{Ellis:2020ena} and Fig.~1 in~\cite{NANOGrav:2023gor} the amplitude and spectral slope of the NANOGrav 15-year data cannot be fitted well by the one-parameter cosmic string model proposed previously. The new data indicate that the spectrum is steeper than what the simple one-parameter cosmic string model can reproduce~\cite{NANOGrav:2023hvm,Antoniadis:2023xlr}.

Cosmic strings can also arise in superstring theory~\cite{Dvali:2003zj,Copeland:2003bj}, as fundamental strings stretched to cosmological sizes. In such models the probability that strings intercommute does not have to be unity as in the field-theory case, and values as low as $p=10^{-3}$ are well motivated~\cite{Jackson:2004zg}. Another possible realisation could be provided by colour flux tubes associated with confinement in pure Yang-Mills theories, which would share similar phenomenological characteristics~\cite{Yamada:2022aax,Yamada:2022imq}. In the simplest approximation, this means multiplying the loop number density by $p^{-1}$ which leads to an identical increase in the GW signal abundance~\cite{Sakellariadou:2004wq,Blanco-Pillado:2017rnf}. Although it has been suggested that this treatment is oversimplified~\cite{Avgoustidis:2005nv}, improvement of this simple result is still a matter of debate~\cite{Auclair:2019wcv} and we adopt it here, allowing the intercomutation probability $p$ to be a free parameter.

{\bf GW spectrum from cosmic strings:} We approximate the evolution of the string network during the expansion of the Universe using the velocity-dependent one-scale~(VOS) model~\cite{Martins:1995tg,Martins:1996jp,Martins:2000cs,Avelino:2012qy,Sousa:2013aaa}.
In this approach, the network is characterised by its correlation length $L$ and string velocity $\bar{v}$.
The evolution of these parameters in a given background is obtained by solving~\cite{Martins:1996jp,Martins:2000cs}:
\begin{eqnarray}
\frac{dL}{dt} &=& (1+\bar{v}^2)\,HL + \frac{\tilde{c}\bar{v}}{2} \ ,
\label{eq:vos1}\\
\frac{d \bar{v}}{dt} &=& (1-\bar{v}^2)
\left[\frac{k(\bar{v})}{L} - 2H\,\bar{v}\right] \ ,
\label{eq:vos}
\end{eqnarray}
where
\begin{equation}
k(\bar{v}) = \frac{2\sqrt{2}}{\pi}(1-\bar{v}^2)(1+2\sqrt{2}\bar{v}^3)
\left(\frac{1-8\bar{v}^6}{1+8\bar{v}^6}\right) \ ,
\end{equation}
$H$ is the Hubble rate and the parameter $\tilde{c} \simeq 0.23$ describes the loop chopping efficiency~\cite{Martins:2000cs}.  

We assume that the only energy-loss mechanism for the network is through the emission of GWs from closed loops that are continuously cut from the long string network as the long strings collide~\cite{Vachaspati:1984gt}. To compute the gravitational wave emission we follow recent Nambu-Goto string simulations~\cite{Blanco-Pillado:2011egf,Blanco-Pillado:2013qja,Blanco-Pillado:2015ana,Blanco-Pillado:2017oxo,Blanco-Pillado:2019vcs,Blanco-Pillado:2019tbi}. These find that the total result is well approximated by just taking into account the population of large loops with initial size $l_i = \alpha_L L \left(t_i\right)$ where $t_i$ is the formation time and $\alpha_L \approx 0.37$, which during an epoch of radiation domination gives $l_i \approx \alpha t_i$ with $\alpha=0.1$ as expected. Smaller loops cut from the network have much higher kinetic energy, which redshifts rather than contributing to the emission. The simulations find that ${\cal F} \sim 0.1$ of the total energy is transferred into large loops and the GW spectrum, and we assume this factor in our calculations. There is one final additional reduction factor $f_r=\sqrt{2}$ that accounts for the energy transferred into peculiar velocities of loops~\cite{Auclair:2019wcv}, which also reduces the total energy contributing to the GW emission.

After a loop is cut from the network it shortens as it loses energy by emitting GWs:
\begin{equation}\label{eq:loopsize}
    l (t) = \alpha_L L \left(t_i \right) - \Gamma G \mu \left(t -t_i \right) \ ,
\end{equation}
where $t_i$ is the loop creation time and $\Gamma \approx 50$ is the total emission power~\cite{Burden:1985md,Blanco-Pillado:2017oxo,Blanco-Pillado:2017rnf}. To compute the spectrum we need to sum the emission over time from all the loops contributing at a given frequency. Starting with the fundamental mode we have~\cite{Avelino:2012qy,Auclair:2019wcv}
\begin{eqnarray}
\label{eq:GWdensity}
\Omega_{\rm GW}^{(1)}(f) &=&
\frac{16\pi}{3 f}
\frac{\mathcal{F}}{f_r}
\frac{G\mu^2 }{H_0^2}
\frac{ \Gamma}{\zeta(q) \alpha_L}
\int_{t_F}^{t_0}\!d\tilde{t}\; n(l,\tilde{t}) \, : \\
n(l,\tilde{t})&=& \frac{\Theta(t_i - t_F)}{\alpha_L \dot{L}(t_i)+\Gamma G\mu}
\frac{\tilde{c} v(t_i)}{L(t_i)^4} 
\bigg[\frac{a(\tilde{t})}{a(t_0)}\bigg]^5
\bigg[\frac{a(t_i)}{a(\tilde{t})}\bigg]^3
\end{eqnarray}
where $t_i$ is the formation time of the contributing loops. This has to be found by solving Eq.~\eqref{eq:loopsize} and using the fact that loops with size $l(\tilde{t},f)=\frac{2}{f}\frac{a(\tilde{t})}{a(t_0)}$ emit at frequency $f$. The $\Theta$ function corresponds to the moment when the network first reaches scaling and begins GW production. The normalisation function $\zeta(q)=\sum_k q^{-q}$ ensures that the power emitted by all modes sums to $\Gamma$ and we have $q=4/3 (5/4)$ for emission mostly through cusps (kinks). 

It is convenient to relate the emission of higher modes to the fundamental mode by using 
 \begin{equation}
\Omega_{\rm GW}^{(k)}(f) =k^{-q}\,\Omega_{\rm GW}^{(1)}(f/k) \, .
 \end{equation}
Summation of higher frequency modes is crucial for correct reproduction of the spectrum, particularly in the case of modified cosmological expansion when the high-frequency part of the spectrum is not a flat plateau~\cite{Cui:2019kkd,Blasi:2020wpy,Gouttenoire:2019kij}. We use a convenient approximation~\cite{Cui:2019kkd}
\begin{equation}
    \label{eq:modesum}
    \Omega_{GW}(f) \simeq \sum_{k=1}^N\Omega_{GW}^{(k)} + \int_{{N+1}}^{\infty}\!dk\;\Omega_{GW}^{(k)} \ ,    
\end{equation} 
which with $N=10^3$ is very accurate while remaining computationally feasible. We find that summing up to $k=10^{12}$ modes is enough to reproduce the high-frequency slopes in the frequency range of interest.

We focus on the emission by cusps setting $q=4/3$. The simulations~\cite{Blanco-Pillado:2011egf,Blanco-Pillado:2013qja,Blanco-Pillado:2015ana,Blanco-Pillado:2017oxo,Blanco-Pillado:2019vcs,Blanco-Pillado:2019tbi} find a more complicated spectrum that is steeper at low $k$ values. However, at large $k$ values the emission is dominated by cusps with $q=4/3$. For spectra falling at high frequencies it is $q$ that dictates the slope, but at large frequencies it is the tail of very high $k$ values that dominates the signal and a pure cusp spectrum is a good approximation.

\begin{figure}
\centering
\includegraphics[width=0.45\textwidth]{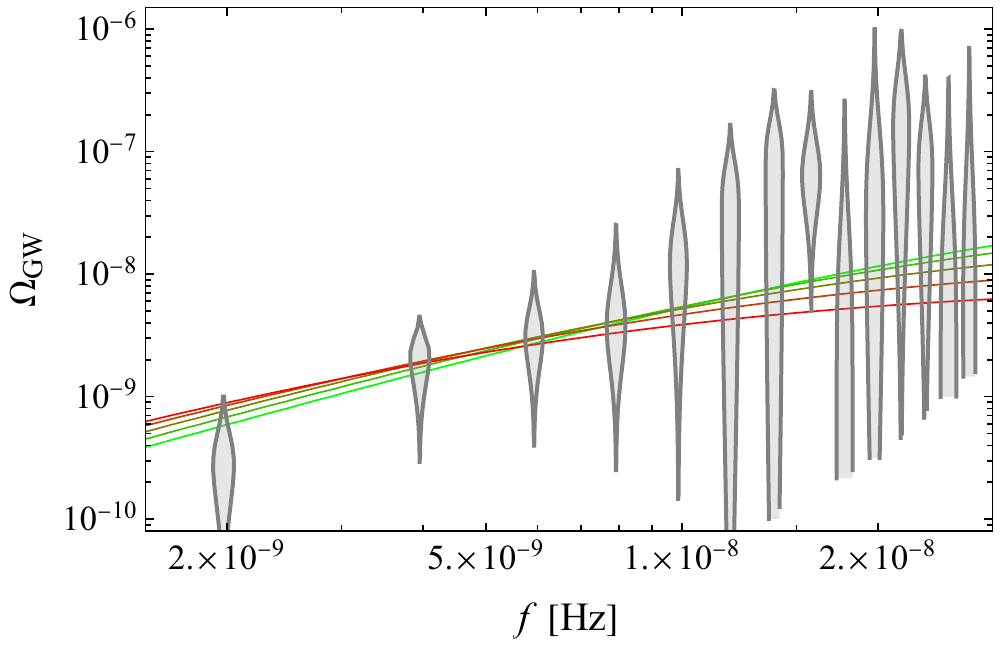}
\caption{\it 
 The curves show the SGWB from cosmic superstrings assuming standard cosmological expansion and the `violins' show the NANOGrav 15-year data. Different curves correspond to choices of $(G\mu, p)$ indicated by points indicated with the same colours in Fig~\ref{fig:SuperCScontours}.}
\label{fig:fit}
\end{figure}

\begin{figure}
\centering
\includegraphics[width=0.45\textwidth]{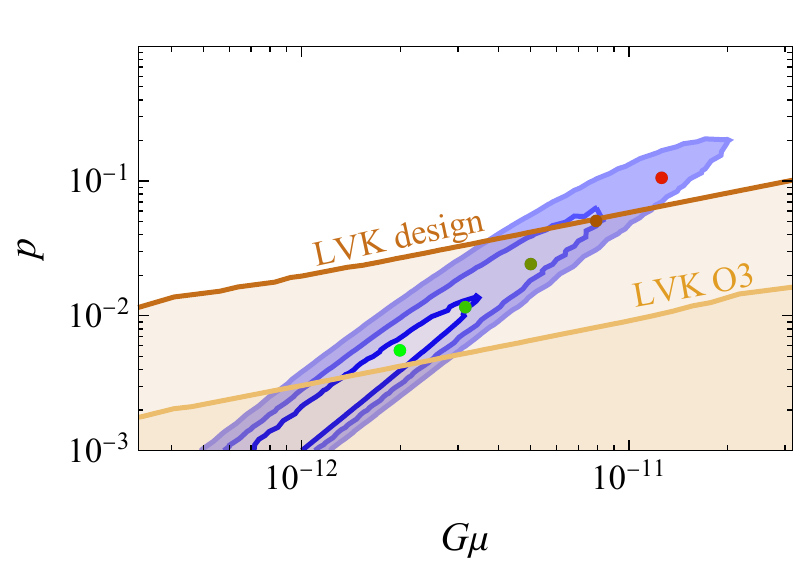}
\caption{\it 
The blue contours correspond to the 68, 95 and 99\% CL ranges for the superstring fit of the NANOGrav 15-year data assuming standard cosmological expansion. The coloured dots correspond to the curves with specific values of the string tension $G\mu$ and intercommutation probability $p$ shown in Figs.~\ref{fig:fit} and \ref{fig:superstringspectra}. The orange regions show the present and projected LVK sensitivities to the SGWB generated by cosmic superstrings.}
\label{fig:SuperCScontours}
\end{figure}

Fig.~\ref{fig:fit} compares the probability density functions of the NANOGrav 15-year data for the GW energy density $\Omega_{\rm GW}$ in different frequency bins, represented as `violins', with the representative choices of the cosmic superstring model parameters $(G\mu, p)$ indicated by the dots of the same colours in Fig.~\ref{fig:SuperCScontours}.
 We see in Fig.~\ref{fig:fit} that the cosmic superstring model can give an acceptable fit to the NANOGrav 15-year data. Fig.~\ref{fig:SuperCScontours} displays the 68, 95 and 99\% CL regions of the cosmic superstring fit of the NANOGrav 15-year data in the $(G\mu, p)$ plane with $p>0.001$. The data prefers $p < 0.1$ at 95\% CL and excludes cosmic strings with $p=1$ as the explanation of the signal by more that 99\% CL (see also Fig. 11 of ref.~\cite{NANOGrav:2023hvm}). Fig.~\ref{fig:SuperCScontours} also shows that most of the 68\% CL region is excluded by the non-observation of a SGWB in the LVK O3 data~\cite{KAGRA:2021kbb}. The LVK design sensitivity will not be able to exclude the cosmic superstring fit to the NANOGrav 15-year data beyond the 95\% CL.
 
{\bf Impact of modified cosmological expansion:}
The cosmic string network evolves and continues to emit GWs throughout the evolution of the Universe. During radiation domination a relatively flat plateau is produced at high frequencies, since the signal redshifts at the same rate as the background, and there are only small changes due to the effects of changes in the number of degrees of freedom. The peak in the spectrum that strings produce at low frequencies is an effect of the recent matter-domination period. This sensitivity makes the signal from cosmic strings a perfect tool to probe the expansion rate in the early Universe~\cite{Cui:2017ufi,Cui:2018rwi,Auclair:2019wcv, Guedes:2018afo, Ramberg:2019dgi, Gouttenoire:2019kij, Gouttenoire:2019rtn,Chang:2019mza,Cui:2019kkd, Gouttenoire:2021jhk}.
We consider now the possibility that the cosmic string signal could explain the NANOGrav observations but remain hidden from other experiments due to non-standard cosmological expansion. We focus on two possible modifications that could diminish the signal, namely an early period of matter domination~\cite{Cui:2017ufi,Cui:2018rwi} and dilution of the network by a period of inflation~\cite{Guedes:2018afo,Cui:2019kkd}.

\begin{figure}
\centering
\includegraphics[width=0.47\textwidth]{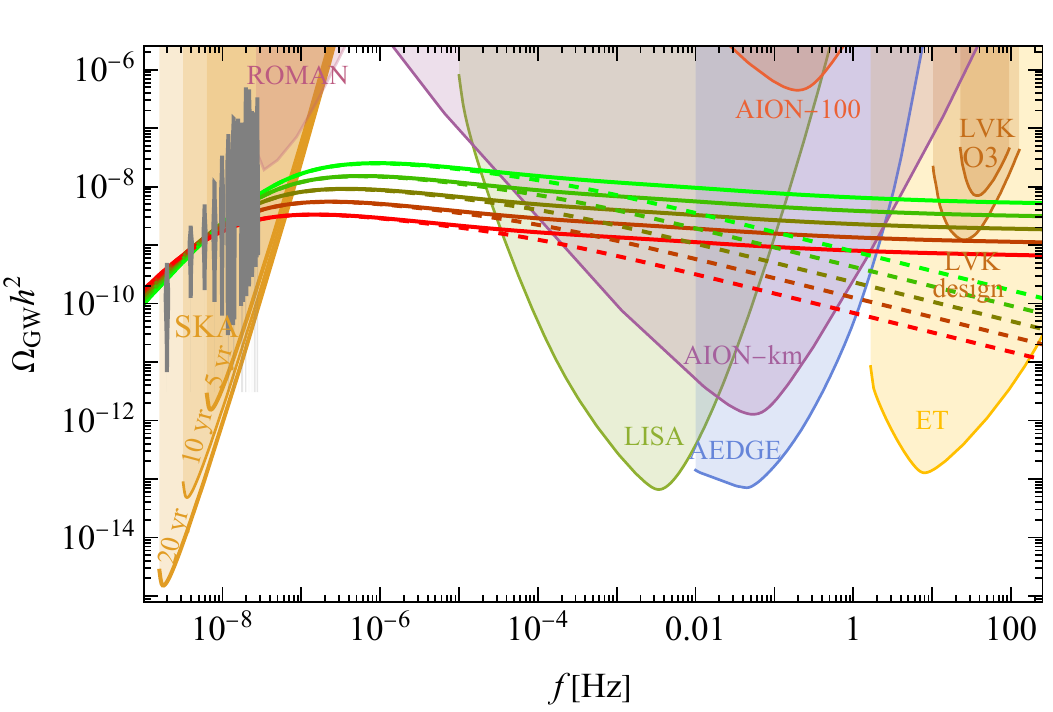}
\caption{\it Extrapolations to higher frequencies of the SGWB 
calculated in the cosmic superstring fits to NANOGrav data assuming
standard cosmological expansion (solid lines) and including the longest possible period
of early matter domination ending at $T=5$ MeV just before BBN (dashed lines). The sensitivities of present and
projected experiments are indicated, including LIGO/Virgo/KAGRA (LVK)~\cite{TheLIGOScientific:2014jea,TheLIGOScientific:2016wyq,LIGOScientific:2019vic}, ET~\cite{Punturo:2010zz,Hild:2010id}, AION~\cite{Badurina:2019hst}, AEDGE~\cite{AEDGE:2019nxb}, LISA~\cite{Bartolo:2016ami,Caprini:2019pxz,LISACosmologyWorkingGroup:2022jok}, the Nancy Roman telescope (ROMAN)~\cite{Wang:2022sxn} and SKA~\cite{Janssen:2014dka}.
\label{fig:superstringspectra}}
\end{figure}

Both these modifications are implemented by modifying the Hubble rate in Eq.~(\ref{eq:vos}) and the resulting expansion in Eq.~(\ref{eq:GWdensity}). We model an early period of matter domination by assuming standard expansion from today back to the moment when the matter domination period ended, and replace the radiation energy density with a matter energy density that redshifts as $a^{-3}$. We assume a long period of matter domination and note that the details of the matter-radiation transition do not affect the signal significantly as long as it is fast. The main constraint is that the transition back to radiation has to finish before Big Bang Nucleosynthesis (BBN) at a temperature of roughly $T_{\rm BBN}=5$ MeV~\cite{Allahverdi:2020bys}, which we use as our benchmark. Fig.~\ref{fig:superstringspectra} shows the impact of this longest possible period of early matter domination on the cosmic string GW spectra through the divergences between the dashed lines and the solid lines obtained for the same string parameters in the case of standard expansion. The modification due to matter domination would enable all the spectra indicated by dots in Fig.~\ref{fig:SuperCScontours} to avoid detection in upcoming runs of LVK and would allow many of the superstring parameter choices currently constrained by the most recent LVK data from their run O3. However, all the spectra modified by matter domination remain within reach of the other indicated future experiments (ET, AION-km, AEDGE and LISA), even though the shapes of the spectra change.~\footnote{We note that a shorter period of matter domination ending at a higher temperature would induce smaller modifications of the spectra that appear only at higher frequencies.}

\begin{figure}
\centering
\includegraphics[width=0.47\textwidth]{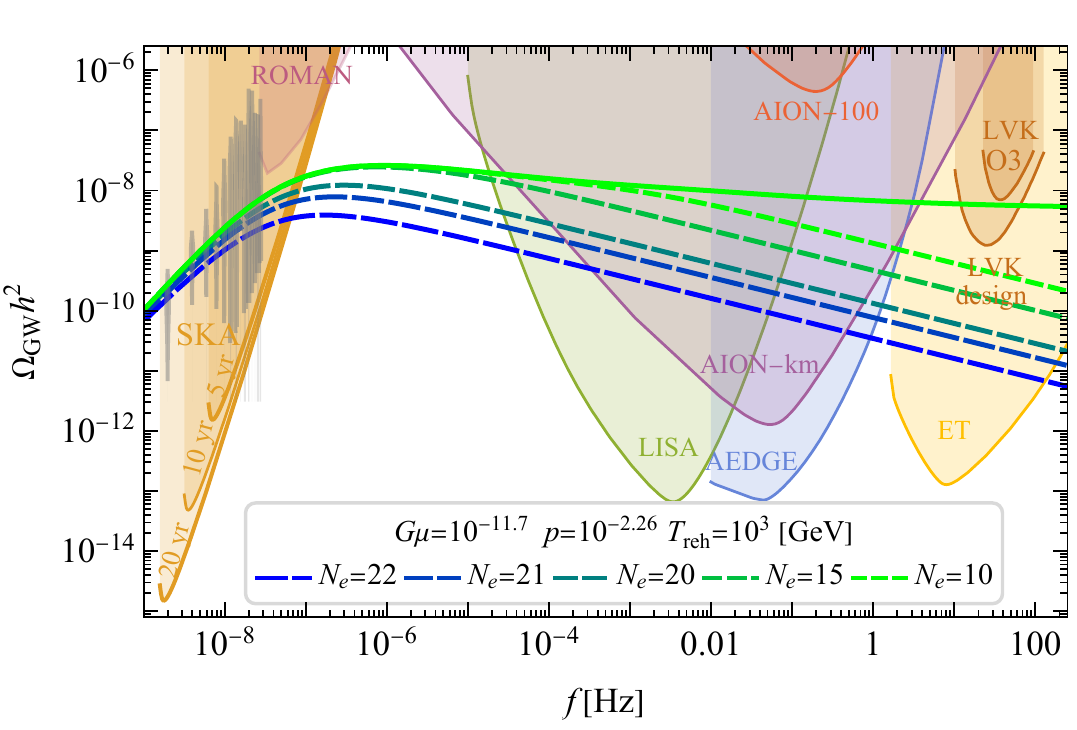}
\includegraphics[width=0.47\textwidth]{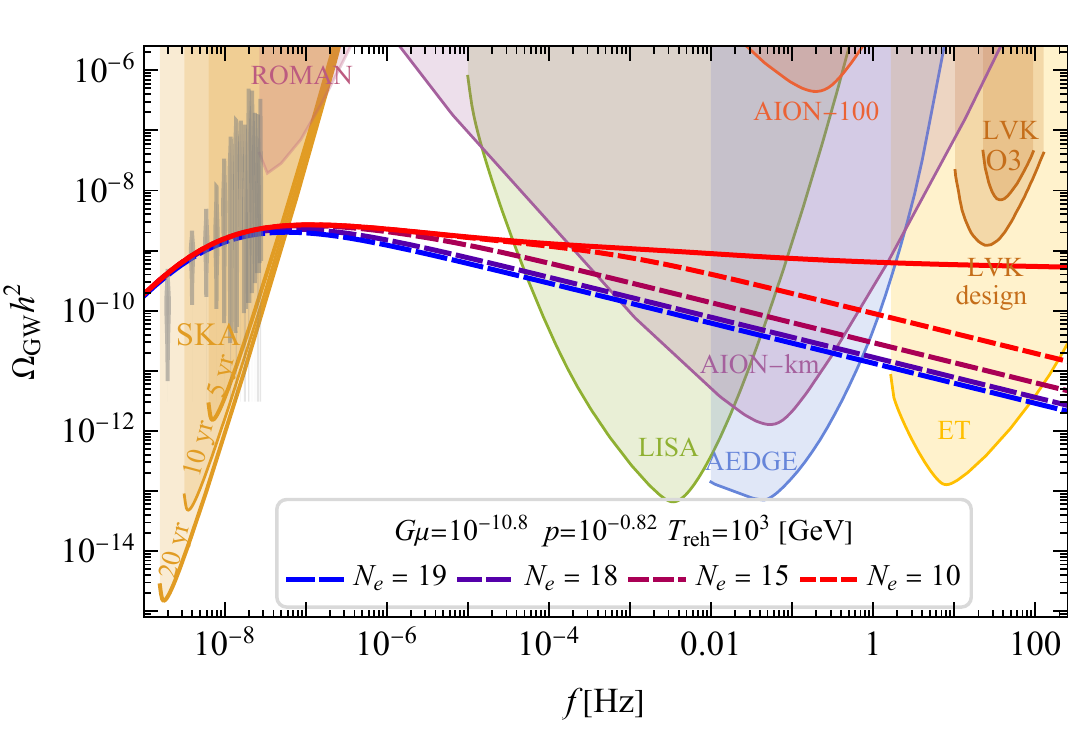}
\caption{\it The impacts of a period of inflation on SGWB signals in 
Fig~\ref{fig:superstringspectra}. Dashed lines show the modification of the spectra by thermal inflation reheating to $T_{\rm reh}=10^3$\,GeV with $N_e$ indicating how long the inflation lasted. The upper panel shows the impact on the strongest signal in 
Fig~\ref{fig:superstringspectra}, shown there in green. Short periods of inflation up to 20 e-folds have little impact on the fit. However, 21 (22) e-folds are excluded by NANOGrav at the 95\% and 99\% CLs, respectively. The lower panel shows the impact on the weakest signal in Fig~\ref{fig:superstringspectra}, shown there in red. Without any modification, this point was already disfavoured at the 99\% CL. Even the shortest period of inflation, corresponding to 19 e-folds, modifies the SGWB signal at low frequencies, reducing the CL to an unacceptable level.
\label{fig:inflation2}}
\end{figure}

We turn next turn to consider the impact of an epoch of inflation. Although standard radiation domination of the Universe must be restored before BBN also in this case, the results are very different from the case of matter domination. This is because the string network is first diluted by inflation and subsequently its abundance increases after reheating as diluted strings redshift slower than radiation so that scaling is reached much later. The frequency at which the spectrum is modified corresponds to the time when strings resume to scaling and not the end of the inflationary period~\cite{Guedes:2018afo,Cui:2019kkd}. The upper and lower panels of Fig.~\ref{fig:inflation2} show the impact of diluting the network by inflation for the strongest and weakest spectrum among those indicated by dots in Fig~\ref{fig:SuperCScontours}. As could be expected, we see that longer periods of inflation result in a longer time before the strings resume scaling. Hence the modification of the spectrum continues to lower frequencies for a larger number of e-folds of inflation. In both cases, we see it is possible to push the modification down to PTA frequencies and the cut-off on e-folds comes when the modification worsens the fit to NANOGrav data beyond $99\%$ CL. In both our examples we see this occurs before the spectra are diminished enough to avoid detection in any of the future experiments and only the reach of LVK can be impacted.

Figure~\ref{fig:inflation2} shows benchmarks for both the reheating temperature, $T_{\rm reh}$, and the number of e-fold, $N_e$. Provided that the inflationary period lasted more than several e-folds all the pre-existing loop population and GWs are redshifted away and the final spectrum is produced only after the strings go back into scaling~\cite{Ferrer:2023uwz}. Thus, for each of our benchmarks, a larger number of e-folds compensated by a higher reheating temperature allowing the network more time to replenish its abundance   would produce the same result. In all cases, the indicated inflationary periods affecting the strings are shorter than the primordial inflation producing the CMB, and strings diluted by primordial inflation would not grow back sufficiently to produce any observable spectra. However, all our results could be reproduced by primordial inflation if the strings were produced a few e-folds after inflation began~\cite{Cui:2019kkd}. On the other hand, if the strings were produced after primordial inflation our results could also be realised by a shorter intermediate period of thermal inflation and, in that case, there could be an additional signal at frequencies corresponding to $T_{\rm reh}$ if that period ended in first-order phase transition~\cite{Ferrer:2023uwz}.  

{\bf Conclusions:} We have reanalysed the recent NANOGrav 15-year data in the context of cosmic strings and superstrings. Cosmic strings are inconsistent with the data, but cosmic super strings are consistent if their intercommutation probability $p \lesssim 0.1$. Our main aim was to study the robustness of the detection prospects of other GW experiments operating at higher frequencies in the context of a possible modification of the expansion rate in the early Universe that would affect the SGWB signal. We find that either an early period of matter domination or some dilution of strings by inflation would preserve some of the points best fitting the NANOGrav data from exclusion by the current and prospective LVK data sets. However, neither of these modifications of standard cosmological expansion can hide the SGWB signal from any of the other planned experiments including ET, AION-km, AEDGE and LISA. Thus, if cosmic superstrings remain a viable source for explaining the NANOGrav and future PTA data these other upcoming experiments will certainly be able to check this possibility despite these possible modifications of the early expansion of the Universe. 

\begin{acknowledgments}
\vspace{5pt}\noindent\emph{Acknowledgments --}
The work of J.E. was supported by the United Kingdom STFC Grants ST/T000759/1 and ST/T00679X/1. The work of M.L. was supported by the Polish National Agency for Academic Exchange within the Polish Returns Programme under agreement PPN/PPO/2020/1/00013/U/00001 and the Polish National Science Centre grant 2018/31/D/ST2/02048. The work of C.L. was supported by the grants 2018/30/Q/ST9/00795 and 2021/42/E/ST9/00260 from the National Science Centre (Poland). The work of V.V. was partially supported by the European Union's Horizon Europe research and innovation program under the Marie Sk\l{}odowska-Curie grant agreement No. 101065736.
\end{acknowledgments}


\bibliography{cosmicstrings}
\end{document}